\documentclass[12pt]{article}

\usepackage{amsmath,amsthm}
\usepackage{amssymb}
\usepackage{comment}
\usepackage{graphicx}
\usepackage{psfrag}
\usepackage{natbib}
\usepackage{hyperref}

\newcommand{\N}{{\operatorname N}}
\newcommand{\LN}{{\operatorname{LN}}}
\newcommand{\E}{{\operatorname E}}

\newcommand{\D}{{\operatorname D}^2}
\newcommand{\Me}{{\operatorname{Me}}}
\newcommand{\EG}{{\operatorname E}_{\textrm{G}}}
\newcommand{\DG}{{\operatorname D}^2_{\textrm{G}}}
\newcommand{\C}{{\operatorname{Cov}}}

\newcommand{\eps}{\varepsilon}
\newcommand{\sG}{\sigma_{\operatorname{G}}}
\newcommand{\sdGh}{\hat{s}_{\operatorname{G}}}
\newcommand{\sdG}{s_{\operatorname{G}}}
\newcommand{\bG}[1]{\bar{#1}_{\operatorname{G}}}
\newcommand{\bA}[1]{\bar{#1}}

\newcommand{\ssG}{s^2_{\operatorname{G}}}
\newcommand{\SSG}{S^2_{\operatorname{G}}}
\newcommand{\ssGh}{\hat{s}^2_{\operatorname{G}}}
\newcommand{\SSGh}{\hat{S}^2_{\operatorname{G}}}

\renewcommand{\)}{\right)}
\renewcommand{\(}{\left(}
\renewcommand{\bar}{\overline}
\newcommand{\FV}{\mathit{FV}}
\newcommand{\PV}{\mathit{PV}}
\newcommand{\const}{\mathrm{const}}
\renewcommand{\tilde}{\widetilde}
\newcommand{\toP}{\stackrel{P}{\longrightarrow}}

\newtheorem{theorem}{Theorem}
\newtheorem{property}{Property}

\begin{document}

\author{Helena Jasiulewicz\footnotemark[1] \footnotemark[2], Wojciech Kordecki\footnotemark[3]}
\title{Additive versus multiplicative parameters:  applications in economics and finance}
\footnotetext[1]{Wroc{\l}aw University of Environmental and Life Sciences, Institute of Economics  and Social Sciences.}
\footnotetext[2]{This work was supported by  National Science Centre, Poland.}
\footnotetext[3]{Faculty of Technical and Economic Science, The Witelon State University of
Applied Sciences in Legnica, Poland.}
\maketitle

\maketitle

\begin{abstract}
In this paper, we pay our attention to multiplicative parameters of random variables
and their estimators. 
We study multiplicative properties of the multiplicative expectation and
multiplicative variation as well as their estimators.  
For distributions having applications in finance and insurance we provide their
multiplicative parameters and their properties. 
We consider, among others, heavy-tailed distributions such as lognormal and Pareto
distributions, applied to the modelling of large losses.
We discuss multiplicative models, in which the geometric mean and
the geometric standard deviation are more natural than their arithmetic
counterparts. 
We provide two examples from the Warsaw Stock Exchange in 1995--2009 and
from a bid of 52-week treasury bills in 1992--2009 in Poland as an illustrative
example.


\end{abstract}

\section{Introduction}
\label{s:intro}

Two  measures frequently used in descriptive
statistics are the arithmetic mean and the standard deviation. 
The geometric mean is used less often, while the geometric standard deviation
connected with the geometric mean is used even more rarely.

When is it better to use arithmetic (additive) parameters and when geometric
(multiplicative) ones? 
A lot of attention has been paid to these problems in the economic and finance
literature. 
One of the firsts papers on this topic was the article by
Latan\'e~\citeyearpar{Latane:Ventures}, who introduced the geometric-mean
investment strategy into the finance and economics
literature.
Weide, Peterson and Maier wrote in their paper \citeyearpar{Weide_et_al}:

\textit{Most of this work has been devoted to the investigation of various
properties of the geometric mean strategy. 
Among the properties of optimal geometric-mean portfolios recently discovered are
(i) they maximize the probability of exceeding a given wealth level in a fixed
amount of time, (ii) they minimize the long-run probability of ruin, and (iii) they
maximize the expected growth rate of wealth.}

In the paper \citep{Weide_et_al}, they consider either the computational
problem of finding the optimal geometric mean portfolio or the question of the
existence of such a portfolio. 
They analysed both of these problems under various assumptions about the investor's
opportunity set and the form of his/her subjective probability distribution of
holding period returns.

Let us assume that the gross return $R$ in a single period has a lognormal
distribution. 
The unknown parameter is $a=\E\(R\)=e^{m+\sigma^2/2}$. 
To estimate this parameter one can use the arithmetic mean of gross returns:
\[
\bar{R}=\frac{1}{N}\sum_{i=1}^N R_i\,.
\]
It is an unbiased estimator of the parameter $a$. 
Another unknown parameter considered in \citep{Cooper} is the geometric mean of
the gross return  $b=\EG\(R\)=e^m$. 
The parameter $b$ can be estimated as the geometric mean
\[
\bG{R}=e^{\bar{\ln R}}=\exp\(\frac{1}{N}\sum_{i=1}^N\ln R_i\).
\]
In \citep{Cooper,Jacquier2003, Jacquier2005}, the expected value $\E\(\bG{R}\)$
is calculated. 
This value is an asymptotically unbiased estimator of $b$. 
Moreover, the variance $\D\(\bG{R}\)$, which tends to zero, is
determined. 
In our paper, we point out that the quality of the geometric estimator should
be examined by the geometric mean and variance, not by their arithmetic
counterparts as in \citep{Cooper,Jacquier2003,Jacquier2005}.

In the paper \citep{Hughson}, the authors point out that forecasting a typical future cumulative return should be more
focused on estimating the median of the future cumulative return than on
the median of the expected cumulative return. 
Expectation of the cumulative return is always higher than the median of the cumulative return.
The probability distribution of returns from risky ventures is positively skewed. 
It is frequently assumed that returns have lognormal distributions. 
For a lognormal distribution, the median and the geometrical
expectation are equal. 
Another distribution frequently used in finance and insurance is the Pareto
distribution, in which the geometric mean is close to the median and far from
the arithmetic mean.

Arithmetic and geometric means are somewhat controversial measurements of the past
and future investment returns.
Critical remarks on this topic are given in the paper
\citep{Missiakoulis_et_al}. 
A review of basic equalities and inequalities in the context of a gross income from
the investment in a discrete time  can be found in the article \citep{Cate}.
 
Properties of various kinds of means can be found in the review paper
\citep{Ostasiewicz:OpRes2000}.

In this paper, unlike in the results discussed above, the issue concerning
multiplicative parameters, including a geometric mean, is also extended with
interpretations and applications of multiplicative variance as a measure of
dispersion. 
Such a measure, as we justify in more detail in the next sections, is
a better and more natural measure of deviation between random variables and
their geometric mean. 

The geometric variance is invariant with respect to multiplication
by a constant. 
From this property it follows that the variance of an economic quantity given in different monetary units is constant, independent of the choice of
the unit. 
For example, if the monetary unit is \$1 or one monetary unit is \$100 then the variance is the same. 
Moreover, the geometric variance is a dimensionless measure of variability. 
For example, it allows to compare the variability of exchange rates between
different currencies.

In Section~\ref{ss:multi} we give definitions and properties of multiplicative
parameters.
We discuss multiplicative models, in which the geometric
mean and the geometric standard deviation are more natural than
their arithmetic counterparts. 
In Section~\ref{ss:distri} we introduce typical
distributions for which the multiplicative parameters are more natural than the
additive ones.
In Section~\ref{ss:estimation} we provide estimators of the multiplicative
parameters considered in Section~\ref{ss:multi} and their properties.
In Section~\ref{s:appli} we give real examples of applications. 
These examples indicate the real benefits of applying the
geometric parameters instead of arithmetic ones in real situations in economics and
finance.

\section{Parameters and models}
\label{s:models}

\subsection{Multiplicative parameters and models}
\label{ss:multi}

Let us define the multiplicative (geometric) mean by
\begin{equation}\label{eq:E_G}
\EG\left(X\right)=e^{\E\left(\ln X\right)},
\end{equation}
where $\Pr\(X>0\)=1$.
From Jensen's inequality it is easy to see that
\begin{equation*}
\EG\(X\)\leq\E\(X\).
\end{equation*}

Below  we give some obvious properties of the geometric mean. 
Equation~\eqref{eq:E_G} implies the formula 
\begin{equation}\label{eq:multi_E}
\EG\(\prod_{i=1}^n X_i\)=\prod_{i=1}^n\EG\(X_i\),
\end{equation}
provided multiplicative expectations of random variables $X_i$ exist. 
In this formula the random variables $X_i$ may be dependent.
Moreover, for every $a>0$
\begin{align}
\notag
\EG\(aX\)&=a\EG\(X\), \\
\intertext{and for every $a\in\mathbb{R}$} 
\label{eq:EG^a}
\EG\(X^a\)&=\(\EG\(X\)\)^a.
\end{align}
From ~\eqref{eq:EG^a} for $a=-1$ we obtain
\begin{equation}\label{eq:multE^-1}
\EG\(\frac{1}{X}\)=\frac{1}{\EG\(X\)}\,.
\end{equation}
Hence, from \eqref{eq:multi_E} and \eqref{eq:multE^-1} we have
\begin{equation*}
\EG\(\frac{X}{Y}\)=\frac{\EG\(X\)}{\EG\(Y\)}.
\end{equation*} 
\begin{property}
If $\EG\(X+Y\)$ exists then
\begin{equation}\label{eq:geo_sumy}
\EG\(X+Y\)\geq \EG\(X\)+\EG\(Y\).
\end{equation}
\end{property}
\begin{proof}
The formula~\eqref{eq:geo_sumy} is by definition equivalent to
\begin{equation}\label{eq:geo_sumy_def}
e^{\E\(\ln\(X+Y\)\)}\geq e^{\E\(\ln\(X\)\)}+e^{\E\(\ln\(Y\)\)}.
\end{equation}
Dividing both sides of \eqref{eq:geo_sumy_def} by $e^{\E\(\ln\(X\)\)}$ we obtain an
equivalent inequality
\[
e^{\E\(\ln\(1+Y/X\)\)}\geq 1+e^{\E\(\ln\(Y/X\)\)}.
\]
Let $T=Y/X$. 
Then, it is sufficient to prove the inequality
\begin{equation*}
e^{\E\(\ln\(1+T\)\)}\geq 1+e^{\E\(\ln T\)}.
\end{equation*}
Let us assume that $T$ is a discrete random variable and $\Pr\(T=x_i\)=p_i$. 
From the inequality~(7.1) from the book~\citep{Mitrinovic:CNI}, p.~6, we
obtain, after the substitution $f\(x\)=\ln\(1+e^x\)$, the inequality
\[
\ln\(\exp\(\sum_{i=1}^n p_i x_i\)+1\)\leq\sum_{i=1}^n p_i\ln\(e^{x_i}+1\).
\]
Substituting $x_i=\ln a_i$ we obtain
\[
\exp\(\sum_{i=1}^n p_i\ln a_i\)+1\leq \exp\(\sum_{i=1}^n p_i\ln\(a_i+1\)\),
\]
which completes the proof of \eqref{eq:geo_sumy} for discrete $X$ and $Y$. 
For any $X$ and $Y$ in the inequality~\eqref{eq:geo_sumy} we approximate $X$
and $Y$ by discrete random variables.
\qed
\end{proof}

The square multiplicative divergence between positive $t$ and 1 is defined by the
following conditions:
\begin{enumerate}
\item\label{enu:f1}
$\displaystyle f\left(t\right)\geq 1$ and
$\displaystyle f\left(1\right)=1$,
\item\label{enu:f2}
$\displaystyle f\left(t\right)=f\left(\frac{1}{t}\right)$,
\item\label{enu:f3}
$f\(t\)$ is an increasing function for $t\geq 1$.
\end{enumerate}
Condition~\ref{enu:f2} means that for any two positive numbers $u$ or $v$:
\[
f\(\frac{u}{v}\)=f\(\frac{v}{u}\).
\]
The function
\begin{equation}\label{eq:geom-dist-log}
f\left(t\right)=e^{\ln^2 t}=t^{\ln t}
\end{equation}
fulfils the above conditions and plays the same role for quotients as $t^2$ for
differences.
It means that $f\(u/v\)$ is a square multiplicative deviation of $u/v$ from 1.

We will define the geometric variance as the multiplicative mean of the
square multiplicative deviation of the random variable $X$ from its geometric
mean:
\begin{equation}\label{eq:V_G}
\DG\left(X\right)
=\EG\(\exp\(\ln^2\frac{X}{\EG\(X\)}\)\)
=e^{\D\left(\ln X\right)}.
\end{equation}
From definition~\eqref{eq:V_G} we have 
\begin{align*}
\DG\(X\)&\geq 1,\\
\DG\(X\)&=1\iff \Pr\(X=\const\)=1.
\end{align*}
The multiplicative (geometric) standard deviation is defined by:
\begin{equation*}
\sG\(X\)=e^{\sqrt{\D\(\ln X\)}}.
\end{equation*}
Note that if $\DG\(X\)\neq 1$ or $\DG\(X\)\neq e$ then $\sG\(x\)\neq\sqrt{\DG\(X\)}$.
A counterpart of 
\begin{equation*}
\sigma\(X\)+\sigma\(Y\)\geq\sigma\(X+Y\)
\end{equation*}
is given by the equation
\begin{equation}\label{eq:G-dysperja-nier}
\sigma_{\operatorname{G}}\(X\)\sigma_{\operatorname{G}}\(Y\)\geq\sigma_{\operatorname{G}}\(X+Y\).
\end{equation} 
However, one cannot compare $\D\(X\)$ and $\DG\(X\)$ because
$\sigma\(X\)=\sqrt{\D\(X\)}$ is represented in the same units as $X$ (e.g.\ in euro
or units of weights or sizes) but $\sG\(X\)$ is dimensionless (may be expressed in
percent after multiplying by 100). 

Apart from function~\eqref{eq:geom-dist-log} 
the function
\begin{equation}\label{eq:SaatyVargas}
f\left(t\right)=e^{\left|\ln t\right|}
\end{equation}
also fulfils the above conditions \citep{SaatyVargas:Dispersion}. 
Note, however, that the function defined by \eqref{eq:SaatyVargas} is a
multiplicative counterpart of $\E\left|X-\E X\right|$, not of the variance
$\D\(X\)$.

Below we give some properties of the multiplicative variance.
Equation~\eqref{eq:V_G} implies the formula 
\begin{equation}\label{eq:multi_D}
\DG\(\prod_{i=1}^n X_i\)=\prod_{i=1}^n\DG\(X_i\),
\end{equation}
provided multiplicative variances of random variables $X_i$ exist
and $X_i$ are independent.
Moreover, for every $a>0$
\begin{align}
\notag
\DG\(aX\)&=\DG\(X\), \\
\intertext{and for every $a\in\mathbb{R}$}
\label{eq:DG^a}
\DG\(X^a\)&=\(\DG\(X\)\)^{a^2}, \\
\notag
\sG\(X^a\)&=\(\sG\(X\)\)^a.
\end{align}
From~\eqref{eq:DG^a} for $a=-1$ we obtain
\begin{equation}\label{eq:multiV^-1}
\DG\(\frac{1}{X}\)=\DG\(X\).
\end{equation}
Hence, if $X$ and $Y$ are independent then from \eqref{eq:multi_D} and
\eqref{eq:multiV^-1} we have
\begin{equation*}
\DG\(\frac{X}{Y}\)=\DG\(X\)\DG\(Y\).
\end{equation*} 

The multiplicative variance and standard deviation are quotient measures of the
deviation between a random variable and its multiplicative mean
$m_{\textrm{G}}=\EG\(X\)$, whereas the additive variance and standard deviation are
difference measures of the deviation between a random variable and its additive mean
$m$. 
Since in the additive case it is useful to define the $k$th interval of the form 
\[
\(m-k\sigma,\;m+k\sigma\),
\]
in the multiplicative case we have the counterpart of the form
\begin{equation}\label{eq:kG-interval}
\(m_{\textrm{G}}\sigma_{\textrm{G}}^{-k},\;m_\textrm{G}\sigma_{\textrm{G}}^k\).
\end{equation} 

Let $\(X,Y\)$ be a two-dimensional random vector. 
We will find the best exponential approximation of a random variable $Y$ by a random
variable $X$. 
To achieve that we will find a multiplicative counterpart of the equation 
\begin{equation*}
\min_{a,b}\E\(Y-\(aX+b\)\)^2=\E\(Y-\(\tilde{a}X+\tilde{b}\)\)^2.
\end{equation*}
The measure of the distance between a random variable $Y$ and  the exponential
function of a random variable $X$ of the form $e^{\alpha X+\beta}$ will be,
according to equation~\eqref{eq:geom-dist-log}, the geometric expectation 
of the random variable $e^{\ln^2 T}$, where 
\[
T=\frac{e^{\alpha X+\beta}}{Y}\,.
\]
Note that
\[
\EG\(e^{\ln^2T}\)=\exp\(\E\ln e^{\ln^2 t}\)
=e^{\E\(\ln^2T\)}=e^{\E\(\ln Y-\(\alpha X+\beta\)\)^2}.
\]
Instead of minimizing the expression $\E\(Y-\(aX+b\)\)^2$ 
we will minimise the expression
\[
\EG e^{\(\alpha X+\beta-\ln Y\)^2}.
\]
Therefore,
\begin{equation*}
\min_{\alpha,\beta}\E\(\ln Y-\(\alpha X+\beta\)\)^2
=\E\(\ln Y-\(\tilde{\alpha}X+\tilde{\beta}\)\)^2,
\end{equation*}
for
\begin{align}
\label{eq:aproks-exp-a}
\tilde{\alpha}&=\frac{\C\(X,\ln Y\)}{\D\(X\)}\,, \\
\label{eq:aproks-exp-b}
\tilde{\beta}&=\E\(\ln Y\)-\frac{\C\(X,\ln Y\)}{\D\(X\)}\E\(X\).
\end{align}
Formulae \eqref{eq:aproks-exp-a} and \eqref{eq:aproks-exp-b} imply that the
function that is the best approximation of the random variable $Y$ has the form
\begin{equation}\label{eq:aproks_exp_1}
y=e^{\tilde{\alpha}\(x-\E\(X\)\)}\EG\(Y\).
\end{equation}
Note that in equation~\eqref{eq:aproks_exp_1} the parameters of the random
variable $X$ are additive whereas the parameters of the random variable
$Y$ are multiplicative. 

The multiplicative econometric model with one explanatory variable is of the form
\begin{equation*}
Y=f\(x\)\eps,
\end{equation*}
where $\eps$ is a random component.
It is frequently  assumed that $\eps$ has a lognormal distribution with parameters
$m$ and $\sigma$. 
Let $Z=\ln Y$. 
Then 
\begin{equation}\label{eq:model_ln}
Z=\ln f\(x\)+\ln\eps
\end{equation}
is an additive model with a random component $\eta=\ln\eps$ with a normal
distribution $\N\(0,\sigma\)$.  
We will denote its trend by $z$, where
\begin{equation}\label{eq:model-m-est}
z=\ln f\(x\).
\end{equation}
An exponential trend is defined by the formula
\begin{equation}\label{eq:regression-mult}
y=f\(x\)=e^{\alpha x+\beta}.
\end{equation}
The trend in the multiplicative model is given by
\begin{equation}\label{eq:trend-m}
y=e^z.
\end{equation}
The behaviour of the variable $y$ in the multiplicative model is reflected by its
geometric mean.

\subsection{Parameters of selected distributions}
\label{ss:distri}

In this section we will determine multiplicative parameters of distributions
frequently applied to the modelling of a finance risk. 
Two heavy-tailed distributions, namely lognormal and Pareto distributions used to
estimate large losses on financial and insurance markets, are especially important.

A random variable $X$ has a lognormal distribution if $Y=\ln X$ has a normal
distribution, $Y\sim\N\(m,\sigma\)$, $\E Y=m$, $\D Y=\sigma^2$. 
Then, the expectation is
\begin{equation*}
\E\left(X\right)=e^{m+\sigma^2/2}
\end{equation*}
and the variance
\begin{equation*}
\D\left(X\right)=e^{2m+\sigma^2}\left(e^{\sigma^2}-1\right).
\end{equation*}
Multiplicative parameters are the following:
\begin{equation*}
\EG\(X\)=\Me\(X\)=e^m=e^{-\sigma^2/2}\E\(X\),
\end{equation*}
\begin{equation*}
\DG\(X\)=e^{\sigma^2},
\end{equation*}
where the median $\Me\(X\) = \EG\(X\)$ and $\DG\(X\)$ depend only on $m$ and $\sigma$, respectively.

The divergence between means $\E\(X\)$ and $\EG\(X\)$ measured by their
relationship $d$ is given by
\begin{equation*}
d\(\sigma\)=\frac{\E\(X\)}{\EG\(X\)}=\frac{e^{m+\sigma^2/2}}{e^m}=e^{\sigma^2/2}
\end{equation*}
and increases exponentially with $\sigma^2$.

In this context, an interesting distribution is the Pareto distribution,
with a cumulative distribution function
\begin{equation}\label{eq:Pareto-cdf}
F_P\(x\)=
\begin{cases}
1-\(\frac{\beta}{x}\)^{\alpha} & \text{for $x\geq\beta$,} \\
0 & \text{for $x<\beta$,}
\end{cases}
\end{equation} 
where $\alpha>0$, $\beta>0$.

The additive parameters of the random variable $X$ are:
\begin{equation*}
\E\(X\)=\frac{\alpha\beta}{\alpha-1}
\end{equation*}
for $\alpha>1$ and
\begin{equation*}
\D\(X\)=\frac{\alpha\beta^2}{\(\alpha-2\)\(\alpha-1\)^2}
\end{equation*}
for $\alpha>2$.

The multiplicative parameters are:
\begin{align}
\label{eq:EG_Pareto}
\EG\(X\)&=\beta e^{1/\alpha}, \\
\label{eq:VG_Pareto}
\DG\(X\)&=e^{1/\alpha^2}
\end{align}
and exist for any $\alpha>0$. 
The median $\Me(X)$ exists for any $\alpha$ and is given by
\begin{equation*}
\Me\(X\)=\beta 2^{1/\alpha}<\EG\(X\).
\end{equation*}
Since
\[
\lim_{\alpha\to\infty} \E\(X\)=\lim_{\alpha\to\infty}\EG\(X\)=1,
\]
for large $\alpha$ we have $\E\(X\)\approx \EG\(X\)$.

\subsection{Estimation of multiplicative parameters}
\label{ss:estimation}

Let us define the following empirical parameters: the geometric mean
\begin{equation}\label{eq:exp_emp_F}
\bG{x}=\(\prod_{i=1}^n x_i\)^{1/n}=\exp\(\frac{1}{n}\sum_{i=1}^n\ln x_i\)
\end{equation} 
and geometric variances
\begin{align}
\label{eq:var_emp_G1}
\ssG&=\(\prod_{i=1}^n\exp\(\ln^2\frac{x_i}{\bG{x}}\)\)^{1/n}
=\exp\(\frac{1}{n}\sum_{i=1}^n\ln^2\frac{x_i}{\bG{x}}\), \\
\label{eq:var_emp_bGh}
\ssGh&=\(\prod_{i=1}^n\exp\(\ln^2\frac{x_i}{\bG{x}}\)\)^{1/\(n-1\)}
=\exp\(\frac{1}{n-1}\sum_{i=1}^n\ln^2\frac{x_i}{\bG{x}}\).
\end{align}
Then, empirical standard deviations are defined as 
\begin{align*}
\ln\sdG&=\sqrt{\frac{1}{n}\sum_{i=1}^n\ln^2\frac{x_i}{\bG{x}}}, \\
\ln\sdGh&=\sqrt{\frac{1}{n-1}\sum_{i=1}^n\ln^2\frac{x_i}{\bG{x}}}.
\end{align*}
Now we can derive from Section~\ref{ss:multi} the equations for estimators of the 
multiplicative parameters and their properties.

Let $X_1,X_2,\dots,X_n$ be a random sample for a population  with cdf $F\(x\)$. 
Let $\theta$ be a multiplicative parameter of $F\(x\)$, e.g.\ $\theta=\EG\(X\)$ or
$\theta=\DG\(X\)$. 
Below we formulate the basic properties of the multiplicative estimators of
such parameters.

The statistic $Z_n=f\(X_1,\dots,X_n\)$ is a multiplicative unbiased estimator
of $\theta$ if $\EG\(Z_n\)=\theta$. 
The $Z_n$ is a multiplicative, asymptotically unbiased estimator of
$\theta$ if 
$\lim_{n\to\infty}\EG\(Z_n\)=\theta$.
The $Z_n$ is a multiplicative consistent estimator of $\theta$ if $Z_n/\theta$
is convergent in probability to 1, denoted as $Z_n/\theta\toP  1$, i.e.
\[
\lim_{n\to\infty}\Pr\(\left|\frac{Z_n}{\theta}- 1\right|>\eps\)=0,
\]
for any $\eps>0$.

\begin{theorem}
Let $X_1,X_2,\dots,X_n$ be a random sample with the multiplicative mean $\EG
X_i=m_{\operatorname{G}}$. 
The statistic $\bG{X}$ is a multiplicative unbiased estimator of
$m_{\operatorname{G}}$.
\end{theorem}
\begin{proof}
From \eqref{eq:EG^a} and \eqref{eq:E_G} we have
\[
\EG\(\bG{X}\)=\EG\(\(\prod_{i=1}^n X_i\)^{1/n}\)
=\(\EG\(\prod_{i=1}^n X_i\)\)^{1/n}
=\(\prod_{i=1}^n\EG\(X_i\)\)^{1/n}.
\]
Then, $\EG\(\bG{X}\)=m_{\operatorname{G}}$.\qed
\end{proof}

Moreover, one can easily calculate the following:
\begin{property}\label{prop:sigma_G_est}
If $X_1,X_2,\dots,X_n$ are independent, identically distributed random
variables and have the multiplicative expectations $m_{\textrm{G}}$ and variances
$\sG^2$ then
\begin{equation*}
\DG\(\bG{X}\)=\(\DG\(X\)\)^{1/n}.
\end{equation*} 
\end{property} 
\begin{proof}
\[
\begin{split}
\DG\(\bG{X}\)
&=\DG\(\prod_{i=1}^n X_i\)^{1/n}
=\(\DG\(\prod_{i=1}^n X_i\)\)^{1/n^2} \\
&=\(\prod_{i=1}^n\DG\(X_i\)\)^{1/n^2}=
\(\DG\(X\)\)^{1/n}=\(\sG^2\)^{1/n}.\hspace{2em}\qed
\end{split}
\]
\end{proof}
Note that $\DG\(\bG{X}\)\to 1$ while $n\to\infty$. 
\begin{theorem}
If $X_1,X_2,\dots,X_n$ are independent, identically distributed random
variables and have the multiplicative expectations $m_{\textrm{G}}$ and variances
$\sG^2$ then
$\bar{X}_G$ is the consistent estimator of $m_G$.
\end{theorem}
\begin{proof}
From the Law of Large Numbers for the sequence $\ln X_1,\ln X_2,\dots,\ln X_n$ we have
\[
\frac{1}{n}\sum_{i=1}^n\ln X_i\toP \E\ln X.
\]
For any continuous $g\(x\)$
\[
g\(\frac{1}{n}\sum_{i=1}^n\ln X_i\)\toP g\(\E\ln X\).
\]
Taking $g\(x\)=e^x$ we have $\bar{X}_G\toP  m_G\,$. 
Hence, $\bar{X}_G$ is the consistent estimator of $m_G\,$.\hspace{2em}\qed
\end{proof}

\begin{theorem}
Let $X_1,X_2,\dots,X_n$ be independent, identically distributed random variables.
The statistic $\SSGh$ is a multiplicative unbiased estimator of $\sG^2$ and $\SSG$
is a multiplicative asymptotically unbiased estimator of $\sG^2$.
\end{theorem}
\begin{proof}
To prove that $\SSGh$ is a multiplicative unbiased estimator of $\sG^2$ we have to
calculate the term
\[
\EG\(\prod_{i=1}^n e^{\ln^2\(X_i/\bG{X}\)}\)^{1/n}.
\]
Let $y_i=\ln x_i$. 
Similarly to proving that  
\[
\hat{S}^2=\frac{1}{n-1}\sum_{i=1}^n\(X_i-\bar{X}\)^2
\]
is an unbiased estimator of $\D\(X\)$ we can prove that $\SSGh$ is a multiplicative
unbiased estimator of $\sG^2$. 
Hence, we omit details. 
As a simple conclusion we obtain that  ${\SSG}$ is a multiplicative asymptotically
unbiased estimator of~$\sG^2\,$.\qed
\end{proof}

\begin{theorem}
Let $X_1,X_2,\dots,X_n$ be independent, identically distributed random
variables.
Then $\SSG$ and $\SSGh$ are the consistent estimators of $\sG^2\,$.
\end{theorem}
\begin{proof}
\[
\SSG=\exp{\frac{1}{n}\sum_{i=1}^n\ln^2\frac{X_i}{\bar{X}_G}}.
\]
Since 
\[
\ln^2\frac{X_i}{\bar{X}_G}=\(\ln X_i-\ln\bar{X}_G\)^2 
=\(\ln X_i\)^2-2\ln X_i\ln\bar{X}_G+\(\ln\bar{X}_G\)^2,
\]
we have
\[
\begin{split}
\frac{1}{n}\sum_{i=1}^n\ln^2\frac{X_i}{\bar{X}_G}
&=\frac{1}{n}\sum_{i=1}^n\(\(\ln x_i\)^2-2\ln X_i\ln\bar{X}_G+\(\ln\bar{X}_G\)^2\) \\
&=\frac{1}{n}\sum_{i=1}^n\(\ln x_i\)^2-2\ln\bar{X}_G
\frac{1}{n}\sum_{i=1}^n\ln X_i+\(\ln\bar{X}_G\)^2. 
\end{split}
\]
From the facts
\[
\begin{split}
&\frac{1}{n}\sum_{i=1}^n\(\ln X_i\)^2\toP \E\(\ln X\)^2, \\
&\frac{1}{n}\sum_{i=1}^n\ln X_i\toP \E\(\ln X\),  \\
&\ln\bar{X}_G\toP \ln m_G\,, \\
&\(\ln\bar{X}_G\)^2\toP \(\ln m_G\)^2,
\end{split}
\]
we obtain by easy calculations
\[
\exp\(\frac{1}{n}\sum_{i=1}^n\ln^2\frac{X_i}{\bar{X}_G}\)\toP 
\exp\(\E\(\ln X- m_G\)^2\)=
\exp\(\D\(\ln X\)\),
\]
which completes the proof. \qed
\end{proof}

Estimators $\hat{\alpha}$ and $\hat{\beta}$ of the parameters $\tilde{\alpha}$ and
$\tilde{\beta}$ given by equations~\eqref{eq:aproks-exp-a} and
\eqref{eq:aproks-exp-b} are given respectively by
\begin{align}
\label{eq:a_mult}
\hat{\alpha}&=\frac{\sum_{i=1}^n x_i\ln y_i-n\bar{x}\,\bar{\ln y}}{\sum_{i=1}^n
x_i^2-n\bar{x}^2}\,, \\
\label{eq:b_mult}
\hat{\beta}&=\bar{\ln y}-\hat{\alpha}\bar{x}.
\end{align}
Estimators of the trend $y$ given by \eqref{eq:regression-mult} has the form
\begin{equation}\label{eq:trend-bm} 
\hat{y}_{\textrm{G}}=\exp\(\hat{z}\),
\end{equation} 
where $\hat{z}=\hat{\alpha}x+\hat{\beta}$.

\section{Applications of the multiplicative model}
\label{s:appli}

Many applications of the geometric mean in economics can be found in the
papers~\citep{Hughson,Jacquier2003}. 
The future portfolio of shares in \citep{Jacquier2003} and the expected gross
return in~\citep{Hughson}
were estimated by the geometric mean.
Cooper in \citeyearpar{Cooper} provided some interesting
considerations on how one can apply the geometric or the arithmetic mean to the
estimation of the discount rate of planned investments.

However, applications nearly always used the multiplicative mean. 
Only in \citep{SaatyVargas:Dispersion} the multiplicative dispersion given by
\eqref{eq:SaatyVargas} was applied, but, as it was explained in
Section~\ref{ss:multi}, that dispersion differs from our standard deviation.

In insurance and finance huge losses are modelled by Pareto or lognormal distributions. 
Such distributions are positively skewed, so their arithmetic expected values are very far from their medians. 
Therefore, the expected values do not reflect the central tendency of these distributions. 
As we will see later, geometric means of distributions do not have such defects. 
Moreover, it is evident that the dispersion around $\EG X$ must be
equal to $\DG X$, not to $\D X$.

Let us only point out that also in other fields of science, multiplicative
parameters give  a better description of some phenomena than additive
ones---see, for example, \citep{NZacharias_et_al:geo} and references
therein.

In the next sections we provide two examples of applications of
multiplicative parameters.
Those examples come from the Polish market and concern the Stock Exchange in Poland.

\subsection{Return index rates}
\label{ss:rates}

Return rates $i_r100\%$ of indexes WIG20 from the Warsaw Stock Exchange in the
years 1995--2009 are given in Table~\ref{tab:indexes},  $r=1995\dots 2009$. 
The accumulation coefficients $a_r=1+i_r$ are given
in the third column.

\begin{table}[!hbt]
\centering
\caption{\label{tab:indexes}Return rates $i_r100\%$ of indexes WIG20 in the years
1995--2009 }
{\renewcommand{\arraystretch}{0.85}

\smallskip
\begin{tabular}{|c|r|c|}
\hline
Year & Rate & Coefficient \\\hline
2009 & $33.47$\% & 1.33 \\
2008 & $-48.21$\% & 0.52 \\
2007 & $5.19$\% & 1.05 \\
2006 & $23.75$\% & 1.24 \\
2005 & $35.42$\% & 1.35 \\
2004 & $24.56$\% & 1.25 \\
2003 & $33.89$\% & 1.34 \\
2002 & $-2.70$\% & 0.97 \\
2001 & $-33.46$\% & 0.67 \\
2000 & $3.40$\% & 1.03 \\
1999 & $43.80$\% & 1.44 \\
1998 & $-16.20$\% & 0.84 \\
1997 & $1.10$\% & 1.01 \\
1996 & $82.10$\% & 1.82 \\
1995 & $8.20$\% & 1.08 \\\hline
\end{tabular}

\smallskip
\textit{Source:}
\verb+ http://www.gpw.pl/analizy_i_statystyki_pelna_wersja+ (November 2014) %
}
\end{table}

The total return at the end of 2009 of an investing initial capital $p=1$ at the
beginning of 1995 (future value $\FV$) is given by the formula:
\begin{equation*}
\FV=\prod_{r=1995}^{2009}a_r\,.
\end{equation*}
Since $\bG{a}=1.0820$,
\begin{equation*}
\FV=\(\bG{a}\)^{15}=3.2656.
\end{equation*}
Using the arithmetic mean $\bA{a}=1.1295$ instead of the geometric mean we
obtain
\begin{equation}\label{eq:capitalA}
\FV'=\(\bA{a}\)^{15}=6.2161,
\end{equation}
which is a two-time overstated estimation of the quantity $\FV$.

Next, we calculate $\sdGh=1.1600$.
Using equation~\eqref{eq:kG-interval} we have the $k$th interval for $\bG{a}$: 
$\(0.9328,1.2550\)$, $\(0.8042,1.4561\)$ and $\(0.6933,1.6890\)$ for $k=1$, $k=2$
and $k=3$, respectively.

If we calculate $\bG{a}=1.1036$ from the 10 years 1995--2004 only, then the
total forecasted return of the capital with the investment of initial capital $p=1$
at the beginning of the year 2005 is equal to $1.6370$. 
The forecast using the arithmetic mean $\bA{a}=1.1447$ from the years 1995--2004 is
equal to $1.9653$. 
The true value of the total return is equal to $1.2185$. 
Therefore, it is more precisely estimated by the geometric mean than by the arithmetic mean.

The analogical conclusion can be drawn from determining the
present value $\PV$ by the geometric and arithmetic means of the discount
factor $v_r=1/a_r$. 
Namely,
\begin{align*}
\PV&=\(\bG{v}\)^{15}=0.3062, \\
\PV'&=\(\bar{v}\)^{15}=0.6504.
\end{align*}

\subsection{The mean annual rate of profitability of treasury bills}
\label{ss:treasury}

A multiplicative model will be used here to describe {the annual market rate with 
investment} for 52-week treasury bills in Poland. 
The use of a multiplicative model can be justified by the fact that
the accumulation of the capital is yielded by the
multiplication, not by the  addition, of gross return from an investment.
Let $R$ denote the annual rate for the 52-week treasury bills and $f\(t\)=ab^t$ be an exponential function of trend.
Assume that (see equation~\eqref{eq:regression-mult})
\begin{equation*}
R=ab^t\eps,
\end{equation*}
where the random component $\eps$ has a lognormal distribution $\LN\(0,\sigma\)$. 

To estimate the unknown parameters $a$ and $b$ (see equations \eqref{eq:a_mult}
and \eqref{eq:b_mult})
of the trend function of the annual rate of interest we make use
of the observations of the average profitabilities from weekly bids in the
years 1992--2009. 
In the observed years, there were from 18 to 56 bids per year. 
For these particular years, the arithmetic and geometric means as well as the medians were taken as the means---see Table~\ref{tab:treasury}.
\begin{table}[!hbt]
\centering
\caption{\label{tab:treasury}Average annual profitabilities of treasury papers in
bids in the years 1992--2009}
{\renewcommand{\arraystretch}{0.85}

\smallskip
\begin{tabular}{|c|c|c|c|c|}
\hline
Year & Arithmetic & Geometric & Median & Number of bids\\\hline\hline
1992&        0.4864&        0.4861&        0.4729&         34\\
1993&        0.3842&        0.3841&        0.3817&         52\\
1994&        0.3238&        0.3231&        0.3816&         52\\
1995&        0.2618&        0.2618&        0.2611&         58\\
1996&        0.2054&        0.2054&        0.2034&         52\\
1997&        0.2210&        0.2210&        0.2193&         56\\
1998&        0.1851&        0.1844&        0.1889&         53\\
1999&        0.1291&        0.1290&        0.1229&         52\\
2000&        0.1761&        0.1761&        0.1780&         49\\
2001&        0.1464&        0.1462&        0.1536&         48\\
2002&        0.0821&        0.0821&        0.0840&         48\\
2003&        0.0536&        0.0536&        0.0549&         48\\
2004&        0.0659&        0.0659&        0.0678&         46\\
2005&        0.0679&        0.0508&        0.0421&         36\\
2006&        0.0420&        0.0419&        0.0421&         23\\
2007&        0.0464&        0.0464&        0.0445&         18\\
2008&        0.0652&        0.0652&        0.0656&         20\\
2009&        0.0465&        0.0465&        0.0475&         49\\\hline
\end{tabular} 

\smallskip
\textit{Source:}
\verb+www.money.pl/pieniadze/bony/archiwum/+ (November 2014)
}
\end{table}

Since differences between them are small, we take as $r_i$ the
arithmetic mean from the annual profitabilities of bids in a particular year.

We will test the hypothesis of normality of $\ln\eps$ using the modified
Jarque--Bera test.
Let $n$ be the sample size, $b_1^{1/2}=m_3/m_2^{3/2}$, $b_2=m_4/m_2^2$, where $m_i$
is the $i$-th central moment of the observations $m_i=\sum\(c_j-\bar{x}\)^i/n$, and
$\bar{x}$ the sample mean.
For testing normality we use the Jarque--Bera test  modified by \cite{Urzua:correct} 
(see also \cite{Thadewald2004Jarque}):
\begin{equation}\label{eq:ALM}
\mathit{ALM}=\frac{\(b_1^{1/2}\)^2}{c_1}+\frac{\(b_2-c_2\)^2}{c_3}.
\end{equation} 
Here the parameters $c_i$, $i=1,2,3$, are given by
\begin{align*}
c_1&=\frac{6\(n-2\)}{\(n+1\)\(n+3\)}\,, \\
c_2&=\frac{3\(n-1\)}{\(n+1\)}\,, \\
c_3&=\frac{24n\(n-2\)\(n-3\)}{\(n+1\)^2\(n+3\)\(n+5\)}\,.
\end{align*}
For our data, we have
$m_2=0.046825093$, $m_3=-0.003081238$, $m_4=0.006135356$, and $n=18$. Hence, we
can calculate that $\textit{ALM}=0.4062$.
The statistic~\eqref{eq:ALM} has an asymptotic $\chi^2$ distribution.
\cite{WurtzKatzgraber:Precise}, using a Monte Carlo simulation, provide precise
quantiles for small samples. 
For the size of sample $n=20$ and the levels $0.01$ and $0.05$ they obtain critical
values $18.643$ and $6.9317$, respectively. 
Therefore, for such critical values one can not reject the null hypothesis of
normality.

\begin{figure}[!hbt]
\centering
\rotatebox{90}{\includegraphics[width=8.6cm]{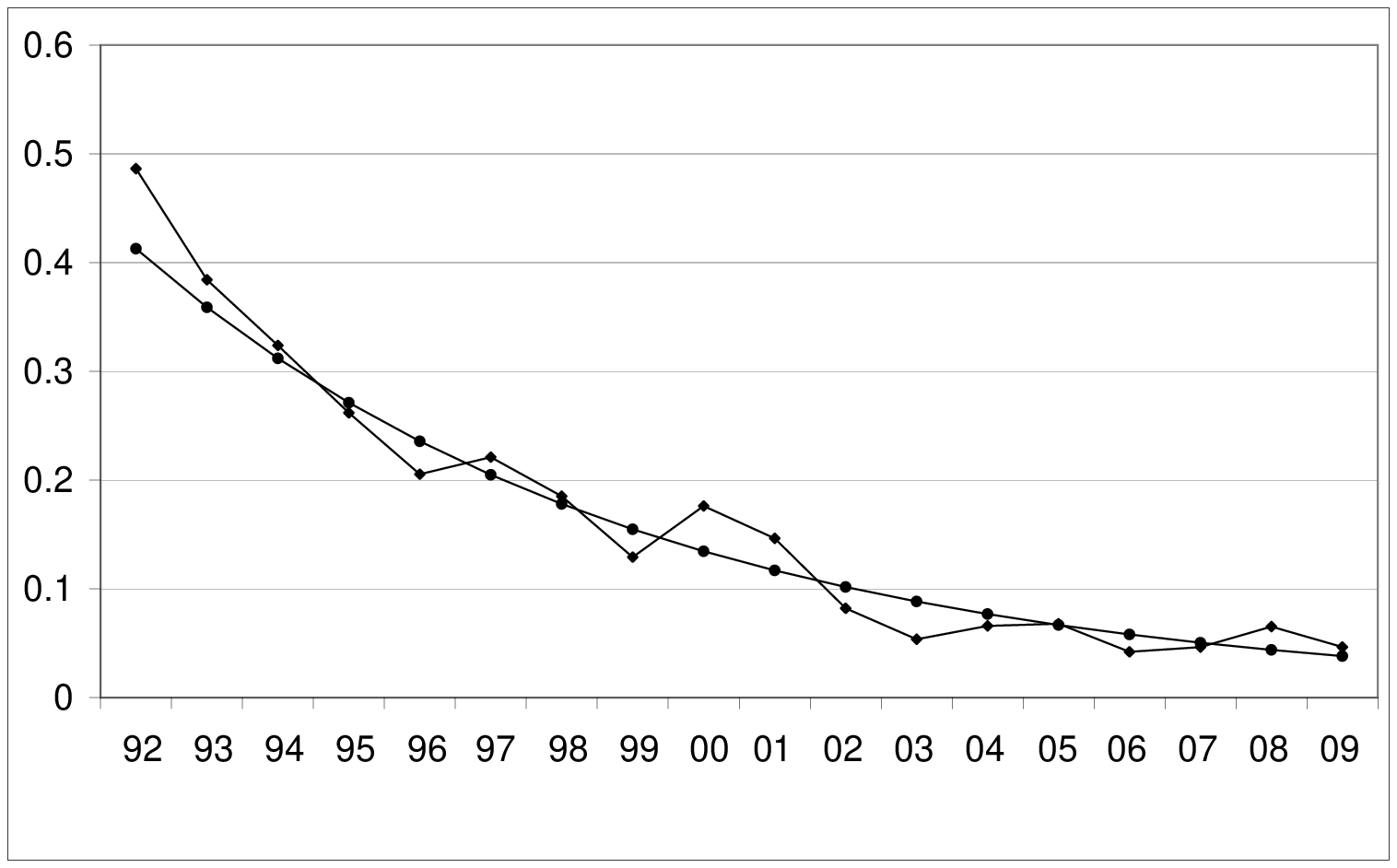}}
\caption{Annual means and their exponential approximations}
\label{fig:treasury}
\end{figure}

Figure~\ref{fig:treasury} shows the average annual profitabilities as well as their exponential approximation 
\[
\hat{r}\(t\)=\exp\(-0.1425t-0.7299\)
\]
given by \eqref{eq:regression-mult}.

The geometric (multiplicative) mean $\bG{r}=0.1245$ was used here to determine the exponential (that is multiplicative) trend of profitability $R$ (see formula~\eqref{eq:aproks_exp_1}). 
For comparison, the arithmetic mean amounts to $\bA{r}=0.1661$, and therefore, since it is significantly greater than $\bG{r}$, it overestimates the long-run returns (see, e.g., \citep{Cooper} and ~\citep{Jacquier2003}). 

\section*{Acknowledgements}
We are grateful to the anonymous reviewer for constructive criticism and to Dr~Cezary Sielu{\.z}ycki for discussions on the final version of the manuscript.



\end{document}